\documentclass[12pt]{article}

 \pdfoutput=1

\usepackage{graphics}
\usepackage{epsfig}

\usepackage{a4wide}
\usepackage{amssymb}
\usepackage{amsmath}
\usepackage{graphicx}
\usepackage{mathdots}
\usepackage{youngtab}
\usepackage{caption}
\usepackage{subcaption}
\usepackage{enumerate}

\newcommand{\be}{\begin{equation}}
\newcommand{\ee}{\end{equation}}
\newcommand{\bea}{\begin{eqnarray}}
\newcommand{\eea}{\end{eqnarray}}

\def\drawbox#1#2{\hrule height#2pt
        \hbox{\vrule width#2pt height#1pt \kern#1pt
              \vrule width#2pt}
              \hrule height#2pt}

\def\Fund#1#2{\vcenter{\vbox{\drawbox{#1}{#2}}}}
\def\Asym#1#2{\vcenter{\vbox{\drawbox{#1}{#2}
              \kern-#2pt       
              \drawbox{#1}{#2}}}}

\def\funda{\Fund{6.5}{0.4}}

\def\symm{\funda\kern-0.4pt\funda}

\def\makeatletter{\catcode`\@=11}
\makeatletter
\def\mathbox#1{\hbox{$\m@th#1$}}%
\def\math@ccstyles#1#2#3#4#5#6#7{{\leavevmode
      \setbox0\mathbox{#6#7}%
      \setbox2\mathbox{#4#5}%
      \dimen@ #3%
      \baselineskip\z@\lineskiplimit#1\lineskip\z@
      \vbox{\ialign{##\crcr
             \hfil \kern #2\box2 \hfil\crcr
             \noalign{\kern\dimen@}%
             \hfil\box0\hfil\crcr}}}}
\def\mathaccstyles{\math@ccstyles\maxdimen}
\def\maththroughstyles{\math@ccstyles{-\maxdimen}}
\def\unity%
 {\maththroughstyles{.45\ht0}\z@\displaystyle {\mathchar"006C}\displaystyle 1}

\begin{document}

\setcounter{table}{0}

\mbox{}
\vspace{2truecm}
\linespread{1.1}

\centerline{\LARGE \bf Discrete $\theta$ and the 5d superconformal index}

\vspace{2truecm}

\centerline{
    {\large \bf Oren Bergman${}^{a}$} \footnote{bergman@physics.technion.ac.il},
    {\large \bf Diego Rodr\'{\i}guez-G\'omez${}^{b}$} \footnote{d.rodriguez.gomez@uniovi.es}
    {\bf and}
    {\large \bf Gabi Zafrir${}^{a}$} \footnote{gabizaf@techunix.technion.ac.il}}

\vspace{1cm}
\centerline{{\it ${}^a$ Department of Physics, Technion, Israel Institute of Technology}} \centerline{{\it Haifa, 32000, Israel}}
\vspace{1cm}
\centerline{{\it ${}^b$ Department of Physics, Universidad de Oviedo}} \centerline{{\it Avda. Calvo Sotelo 18, 33007, Oviedo, Spain }}
\vspace{1cm}

\centerline{\bf ABSTRACT}
\vspace{1truecm}

5d Yang-Mills theory with an $Sp(N)$ gauge group admits a discrete analog of the $\theta$ parameter.
We describe the origin of this parameter in ${\cal N}=1$ theories from Type I' string theory,
and study its effect on the 5d superconformal fixed point theories with
an $Sp(1)=SU(2)$ gauge group by computing the superconformal index.
Our result confirms the lack of global symmetry enhancement in the so-called $\tilde{E}_1$ theory.

\noindent

\newpage

\tableofcontents

\section{Introduction}

Interacting quantum field theories in 5d are non-renormalizable and therefore
do not generically exist as microscopic theories.
Nevertheless there is compelling evidence that there exist strongly-interacting ${\cal N}=1$ supersymmetric 
fixed point theories in 5d, some of which have relevant deformations corresponding to ordinary 
gauge theories with matter \cite{Seiberg:1996bd,Morrison:1996xf,Douglas:1996xp,Intriligator:1997pq,Aharony:1997ju}.

The simplest set of examples has a gauge group $SU(2)$ and $N_f \leq 7$ fundamental hypermultiplets \cite{Seiberg:1996bd}.
This set of fixed point theories is particularly interesting since it was argued to exhibit an exotic global 
symmetry $E_{N_f+1}$.
This is not visible in the gauge theory action, which exhibits only an $SO(2N_f) \times U(1)_T$ global symmetry,
where $U(1)_T$ is the topological symmetry
associated to the conserved current $j_T = *\mbox{Tr}(F\wedge F)$.
It can however be inferred by a particular string theory embedding of the gauge theory 
using a D4-brane in Type I' string theory.
When the D4-brane coincides with the O8-plane and $N_f$ D8-branes at one of the boundaries, 
the low energy supersymmetric gauge theory
has an $Sp(1)=SU(2)$ gauge symmetry, 
and $N_f$ matter multiplets in the fundamental representation.
The fixed point theory corresponds to the limit where the dilaton blows up locally at this boundary,
which is possible only for $N_f\leq 7$.
In particular, this explains the enhancement of the global symmetry to $E_{N_f+1}$, as a result
of the enhancement of the 9d gauge symmetry on the D8-branes due to massless D0-branes
\cite{Polchinski:1995df,Matalliotakis:1997qe,Bergman:1997py}.

The enhanced global symmetry has recently been confirmed by an impressive calculation 
of the superconformal index, including instanton corrections, for the $SU(2)$ theory with $N_f\leq 5$ \cite{KKL}.
The index exhibits explicitly the conserved current multiplets associated with the $E_{N_f+1}$ symmetry.
The extra currents not contained in $SO(2N_f)$ correspond to instanton-particles.

For $N_f=0$ it was argued in \cite{Morrison:1996xf} that there exists a second fixed point theory,
the so-called $\tilde{E}_1$ theory, in which the $U(1)_T$ global symmetry is not enhanced.
It was argued that this theory is associated with a discrete analog of the $\theta$ parameter in 5d \cite{Douglas:1996xp}.

In this note we will revisit the $\tilde{E}_1$ theory. 
We will explain how to embed it in the Type I' description by a previously overlooked discrete choice in the background.
We will then compute the superconformal index for this theory by adapting the computation of \cite{KKL}
to this discrete choice, and exhibit the lack of symmetry enhancement.
We will also offer an alternative computation of the index, treating $SU(2)$ as $SU(N)$ with $N=2$.
In this approach the $\tilde{E}_1$ theory corresponds to the theory with a CS term at level $\kappa=1$.
Along the way we also clarify how the $SU(2)$ theory with CS level $\kappa=2$ is equivalent to that with $\kappa=0$.
We conclude by mentioning some future directions and raising some questions about the string theory interpretation.

\section{The $\tilde{E}_1$ theory}

In \cite{Morrison:1996xf} Morrison and Seiberg showed that there are two more interacting 
fixed points that can be reached by relevant deformations
of the $E_{N_f+1}$ theories.
Starting with the $E_2$ theory, there is a
two-parameter space of relevant deformations spanned by
the bare coupling $t_0 \equiv 1/g_0^2$ and the flavor mass $m$.
These can be thought of as VEVs of scalars in background vector multiplets associated to
the global $E_2 = SU(2)\times U(1)$ symmetry. In particular $m$ is associated to the $U(1)$ part, 
and the combination $m_0\equiv t_0-2m$ to the $SU(2)$ part.
The origin of the $(m_0,m)$ plane is the $E_2$ theory.
Turning on $m>0$ with $m_0=0$, one flows to the $E_1$ theory with $E_1=SU(2)$
global symmetry. On the other hand for $m_0>0$ and $m=0$ one flows to the
the free $D_1$ theory with $SO(2)\times U(1)$ global symmetry.
For $m_0>0$ and $m<0$ there is a singular locus where $m_0 + 4m = 0$, along which the effective
coupling diverges. This suggests that in this direction one reaches a new interacting
fixed point with only a $U(1)$ global symmetry, the so-called $\tilde{E}_1$ theory.
This theory has only one relevant parameter $s=m_0 + 4m$.
For $s>0$ it flows back to the free $D_1$ theory, but for $s<0$ it flows to another
interacting fixed point without any global symmetry, the $E_0$ theory.

A similar conclusion was reached by considering the geometric realization of the $E_{N_f+1}$
fixed points as singular CY spaces with collapsed del-Pezzo surfaces \cite{Morrison:1996xf,Douglas:1996xp}.
The flows described above correspond to shrinking a 2-cycle inside the del-Pezzo surface
and then blowing one up in the CY space transverse to the del-Pezzo surface,
{\em i.e.} a flop transition.
For the surface describing the $E_2$ theory, $B_2$, there are two inequivalent choices
leading either to the Hirzebruch surface $B_1=F_1$ or to the direct product $CP^1\times CP^1$.
The latter corresponds to the $E_1$ theory and the former to the $\tilde{E}_1$ theory.

\begin{figure}[htbp]
\begin{center}
\includegraphics[scale=0.50]{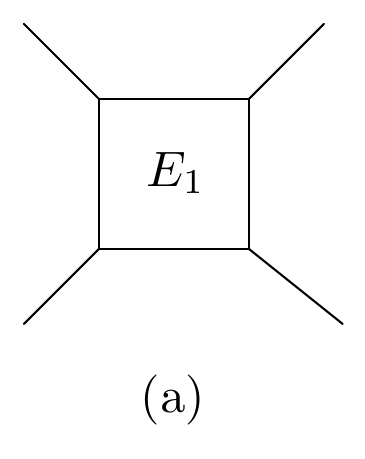}
\hskip 0.5cm
\includegraphics[scale=0.50]{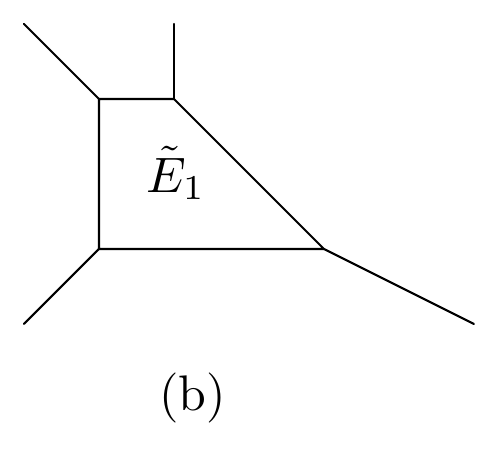}
\hskip 0.5cm
\includegraphics[scale=0.50]{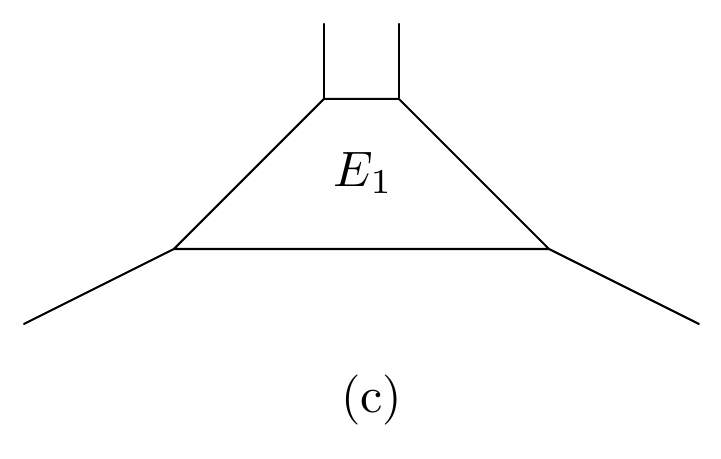}
\caption{Pure $SU(2)$: (a) $E_1$ theory, (b) $\tilde{E}_1$ theory, (c) another representation of $E_1$.
The webs are drawn with the assumption that $g_s=1$ and $C_0 = 0$.}
\label{SU(2)_webs}
\end{center}
\end{figure}

\begin{figure}[htbp]
\begin{center}
\includegraphics[scale=0.50]{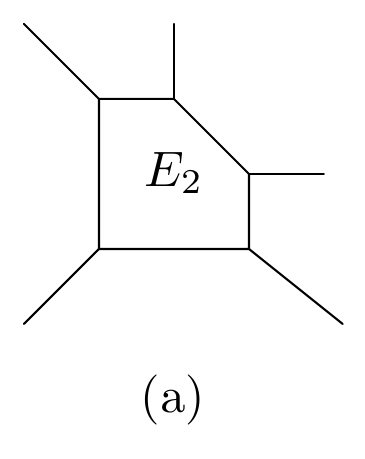}
\hskip 0.4cm
\includegraphics[scale=0.50]{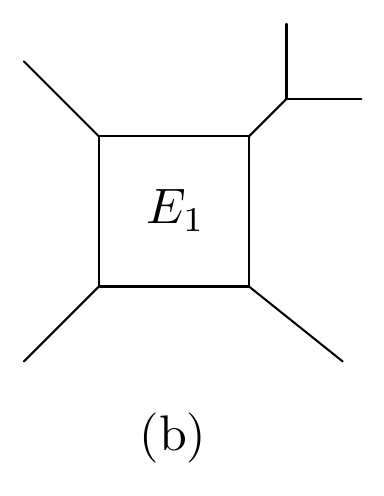}
\hskip 0.4cm
\includegraphics[scale=0.50]{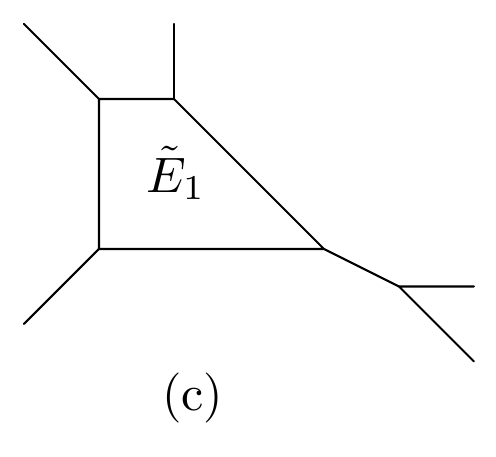}
\hskip 0.4cm
\includegraphics[scale=0.50]{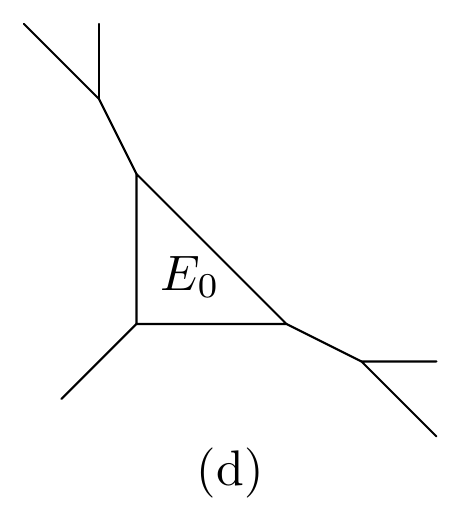}
\caption{$SU(2)$ with one flavor: (a) $E_2$ theory, (b) positive mass deformation, (c) negative mass deformation,
(d) $E_0$ theory.}
\label{E2_web}
\end{center}
\end{figure}

Related to this description is the description of the 5d fixed point theories in terms of Type IIB 5-brane 
webs \cite{Aharony:1997ju}. In this picture 5d gauge theories are constructed using 
a configuration of 5-branes in Type IIB string theory, in which the gauge fields live
on the worldvolumes of D5-branes that are suspended between NS5-branes.
The web-like configuration is supported by semi-infinite $(p,q)$5-branes carrying the appropriate
charges for charge conservation.
The webs for the $E_1$ and $\tilde{E}_1$ theories are shown in Fig.~\ref{SU(2)_webs}.
Flavors can be added by attaching semi-infinite D5-branes.
In particular with one flavor, it is easy to generate the flow from the $E_2$ theory to either
the $E_1$ or $\tilde{E}_1$ theory, Fig.~\ref{E2_web}(a,b,c).
The $E_0$ theory can be reached by deforming the $\tilde{E}_1$ web as in Fig.~\ref{E2_web}(d).

Since the $E_1$ and $\tilde{E}_1$ theories are obtained by deforming with opposite signs for
the flavor mass $m$, it was argued that the difference between the two gauge theories
is a 5d analog of the $\theta$ parameter of Yang-Mills theory \cite{Douglas:1996xp}.
Indeed, 5d $SU(2)$ gauge theory, and more generally $Sp(N)$ gauge theory, admits a $\mathbb{Z}_2$-valued 
$\theta$ parameter associated with $\pi_4(SU(2)) = \mathbb{Z}_2$.
This is a discrete 5d analog of the familiar 4d $\theta$ parameter associated with $\pi_3(SU(2))=\mathbb{Z}$.
In 4d, Euclidean gauge field configurations with instanton number $n\in \pi_3(SU(2))$ are weighted
by a phase $e^{in \theta}$. 
The non-trivial element of $\pi_4(SU(2))$ is associated to a $\mathbb{Z}_2$-charged instanton 
in 5d.\footnote{This is related to the 4d global anomaly of \cite{Witten:1982fp}.
This anomaly was observed by constructing a non-trivial path in configuration space 
that interpolates between a 4d gauge field configuration and its global transformation by an
$SU(2)$ element in the non-trivial class of $\pi_4(SU(2))$.
Along such a path an odd number of eigenvalues of the Dirac operator change sign,
so there is an anomaly if the are an odd number of fermions in the fundamental representation.
This path can be thought of as a $\mathbb{Z}_2$-valued instanton in 5d.}
The 5d path integral thus has two contributions, and we have a choice of taking the sum or difference
of the two. This is interpreted as the discrete choice of the $\theta$ parameter.
When massive flavors are present, this parameter can be absorbed into the sign of their mass. 
Therefore, if there is a massless flavor, $\theta$ is physically irrelevant. 
There is only one theory with one flavor, the $E_2$ fixed point.
But in deforming it by giving mass to the flavor there are two choices, leading to
the $E_1$ and $\tilde{E}_1$ theories.

\section{Type I' string description}

One can now ask how the discrete $\theta$ parameter of the 5d $Sp(N)$ gauge theory is realized in the Type I' description.
This should correspond to some discrete choice that one can make in the Type I' background.
Indeed such a choice exists. 
In 10d Type IIA string theory there is a one-form gauge field in the RR sector, $C_1$.
In reducing to 9d by compactifying on a circle this gives rise to a 9d 
$\theta$ parameter, $\theta_9 = \int_{S^1} C_1$. 
Type I' string theory involves a projection by the combination of spatial and worldsheet reflections,
under which $\theta_9$ is odd, and therefore projected out.
However, since $\theta_9 \sim \theta_9 + 2\pi$, there is a discrete remnant corresponding to
the choice of $\theta_9 = 0$ or $\pi$.\footnote{A similar choice exists in Type I string theory,
implying the existence of a new 10d string \cite{Sethi:2013hra}.}

Associated to $\theta_9$ is a non-BPS D(-1)-brane \cite{Bergman:1999ta}
whose contribution to the path integral comes with a phase $e^{i\theta_9}$.
(The D(-1)-brane is related by 
T-duality to the non-BPS D0-brane 
of Type I string theory \cite{Sen:1998ki,Witten:1998cd}.)
The D(-1)-brane can be regarded as a discrete ``gauge instanton" in the 9d $O(2N_f)$ theory on the 
D8-branes. Its charge is associated with the non-trivial element of $\pi_8(O) = \mathbb{Z}_2$.
Alternatively, it can also be regarded as an instanton in the 5d $Sp(N)$ theory on the D4-branes,
or as an instanton in the 1d $O(k)$ theory on a collection of $k$ D0-branes.
Mathematically, this is the statement of Bott periodicity, which relates
\be
\pi_8(O) = \pi_4(Sp) = \pi_0(O) \,.
\ee
The latter is simply a Wilson line in Euclidean time, where the $O(k)$ gauge field undergoes
a gauge transformation in the negative-determinant component of $O(k)$.
(This is related by T-duality to the realization of the Type I D0-brane as a Wilson line in the Type I D1-brane 
\cite{Sen:1998tt}.)
We are therefore lead to identify the $\theta$ parameters in the different dimensions:
\be
\theta_9 = \theta_5 = \theta_1\,. 
\ee

\begin{figure}[htbp]
\begin{center}
\includegraphics[scale=0.50]{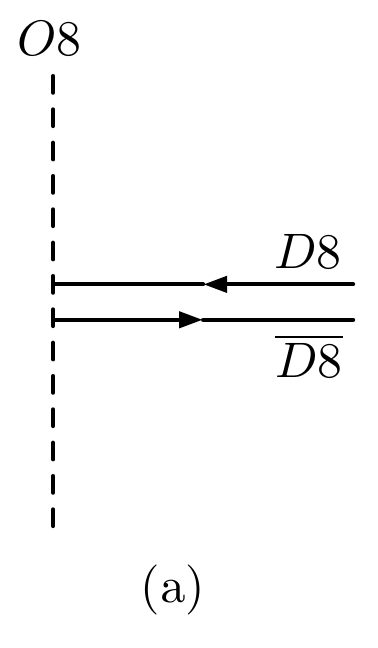}
\hskip 1cm
\includegraphics[scale=0.50]{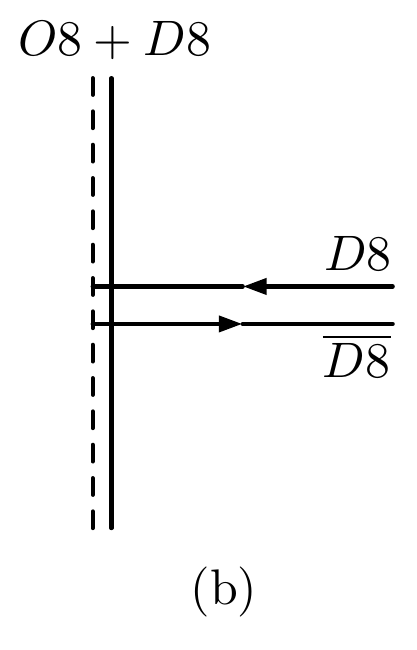}
\hskip 1cm
\includegraphics[scale=0.50]{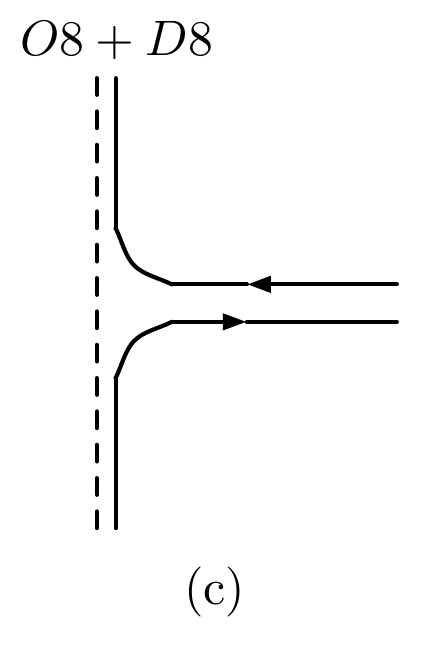}
\caption{Type I' description of the $\theta$ parameter: (a) A transverse D8-$\overline{\mbox{D8}}$ forms a $\theta$ domain wall,
(b) adding a flavor D8-brane leads to an instability, (c) the branes break and reconnect, eliminating the domain wall.}
\label{O8_D8}
\end{center}
\end{figure}

It is useful to construct a background in which $\theta_9$ changes between its two possible values.
This is achieved by considering the 9d ``magnetic" dual of the non-BPS D(-1)-brane,
which is a non-BPS D7-brane. Depending on the size of the interval in the Type I' background,
this corresponds to either a D7-brane localized on the O8-plane, or to a D8-brane-anti-D8-brane
combination stretched along the interval, Fig.~\ref{O8_D8}(a) \cite{Bergman:1999ta}.
(These are T-dual to the D8-brane and D7-brane of Type I string theory \cite{Witten:1998cd}.)
Either way, this creates a domain wall in 9d across which $\theta_9$ changes from 0 to $\pi$.
So when the D(-1)-brane crosses the D7-brane it acquires a $-1$ phase.
This also means that the 5d
$\theta$ parameter jumps across the wall.
In the absence of flavors, {\em i.e.}, no D8-branes transverse to the interval, this creates a clear
separation between two regions in 5d with different values of $\theta$, corresponding to the two distinct vacua
of the 5d $Sp(N)$ gauge theory.
When a flavor D8-brane is added the configuration becomes unstable; there is a tachyon at the
intersection of the D8-brane and the D7-brane, leading to the absorption of the D7-brane by the D8-brane. 
In the stretched D8-$\overline{\mbox D8}$ description of the non-BPS D7-brane the branes break and reconnect, as shown
in Fig.~\ref{O8_D8}(c). 
In other words, the domain wall disappears, and the two vacua are physically equivalent.
This is exactly what we expect from the point of view of the 5d gauge theory.
The $\theta$ parameter can be transformed away in the presence of massless fermion matter.

\section{Superconformal index}

The superconformal index is a characteristic of superconformal field theories that essentially
counts a class of BPS operators
\cite{Kinney:2005ej}. As such, it is protected from continuous deformations of the theory, and can be
used as a diagnostic of non-perturbative physics like duality and enhanced symmetries.
Given a supercharge $Q$ and its conjugate in radial quantization $S$, 
the associated superconformal index is defined in general by
\be
I(\mu_i) = \mbox{Tr}\left[(-1)^F e^{-\beta\Delta} e^{\mu_i q_i}\right] \,,
\ee
where $\Delta = \{Q,S\}$ is the Hamiltonian in radial quantization, $\mu_i$ are chemical potentials associated
to symmetries that commute with $Q$, and $q_i$ are the corresponding charges.
The BPS states contributing non-trivially to $I$ satisfy $\Delta = 0$.
The above expression can be translated to a Euclidean functional integral in the field theory on $S^{d-1}\times S^1$
with appropriately twisted periodicity conditions on $S^1$.

The computation is greatly simplified using supersymmteric localization, which reduces the functional integral
to ordinary matrix integrals.
This was done for 5d ${\cal N}=1$ theories in \cite{KKL}, leading to an expression that factorizes into
a 1-loop determinant contribution and a contribution of ``instanton particles"\footnote{We will refer to these
as instanton particles to differentiate them from the gauge instantons discussed previously.
The latter are true instantons in the sense of being pointlike in 5d, whereas the former
are particles with world-lines in 5d.}:
\be
I(x,y,m_i,q) = \int [{\cal D}\alpha]\, I_{loop}(\alpha,x,y,m_i) \, |I_{inst}(\alpha,x,y,m_i,q)|^2 \,.
\label{eq:index}
\ee
The integral is taken over the holonomy matrix $\alpha$ and includes the Haar measure of the gauge group.
The other parameters are fugacities associated with the Cartan generators of the global symmetry
$SO(5)\times SU(2)_R \times SO(2N_f) \times U(1)_T$. 

For each type of superfield,
the 1-loop contribution is given by a Plethystic exponential of the one-letter index
\be
I_{loop} = PE[f(\cdot)] = {\rm exp}\left[\sum^{\infty}_{n=1} \frac{1}{n} f(\cdot^n)\right] \,.
\label{eq:plesh} 
\ee 
The vector multiplet contributes
\be
f_{vector}(x,y,\alpha) = - \frac{x (y + \frac{1}{y})}{(1 - x y)(1 - \frac{x}{y})}
\sum_{\bold{r}\in \bold{R}}e^{-i\bold{r}\cdot\alpha} \,,\label{eq:vec}
\ee
where $\bold{R}$ is the root lattice,
and each matter multiplet contributes 
\be
f_{matter}(x,y,m_i,\alpha) = \frac{x}{(1 - x y)(1 - \frac{x}{y})}\sum_{\bold{w}\in \bold{W}}\sum^{N_f}_{i=1} 
\left(e^{i\bold{w}\cdot\alpha+im_i}+e^{-i\bold{w}\cdot\alpha-i m_i}\right) \label{eq:mat} \,,
\ee
where the sum is over the appropriate weights in the weight lattice.

The instanton contribution is given by a product of a contribution from instanton particles located
at the south pole of the $S^4$ and a contribution of anti-instanton particles located at the north pole.
Each is 
expressed as a power series in the instanton number,
\be
\label{instanton_sum}
I_{inst}(\alpha,x,y,m_i,q) = 1 + q Z_1(\alpha,x,y,m_i) + q^2 Z_2(\alpha,x,y,m_i) + \cdots \,,
\ee
where $Z_k$ is the 5d Nekrasov $k$-instanton partition function \cite{Nekrasov:2002qd}.
The computation of $Z_k$ involves following the ADHM procedure for quantizing the multi-instanton moduli space,
described by supersymmetric gauge quantum mechanics with a dual gauge group.
This boils down to a contour integral over the Cartan subalgebra of the dual gauge group.
The result depends on both the gauge group and the matter content. Exact results for $SU(N)$,
$Sp(N)$ and $SO(N)$ with various matter were obtained in \cite{Nekrasov:2004vw}.

\subsection{Pure $SU(2)$}

For $SU(2)=Sp(1)$ the dual group for $k$ instantons is $O(k)$.
This is of course the gauge symmetry of the D0-brane theory in the Type I' string theory description.
This group has two disconnected components denoted $O(k)_+ = SO(k)$ and $O(k)_-$.
The latter is the set of determinant $-1$ elements of $O(k)$ and does not form a group.
For $k=2n$, the torus action of the group is generated by
$\mbox{diag}(e^{i\sigma_2\phi_1},\ldots,e^{i\sigma_2\phi_n})$ for $O(2n)_+$
and by $\mbox{diag}(e^{i\sigma_2\phi_1},\ldots,e^{i\sigma_2\phi_{n-1}},\sigma_3)$ for $O(2n)_-$.
In other words the dimension of the Cartan subalgebra of $O(2n)_-$ is smaller by 1 than that of $O(2n)_+$.
For $k=2n+1$ the torus action is generated by $\mbox{diag}(e^{i\sigma_2\phi_1},\ldots,e^{i\sigma_2\phi_n},\pm 1)$.
The $k$-instanton index has two contributions coming from the the two components $O(k)_+$ and $O(k)_-$,
corresponding, respectively, to the sectors of the supersymmetric quantum mechanics without and with the discrete Wilson line.

The general expressions for the two contributions, $Z_k^+$ and $Z_k^-$, were given in \cite{KKL}.
Let us reproduce here just the 1-instanton functions.
The $O(1)_+$ part contributes
\be
Z^{+}_{1} = \frac{x^2}{(1 - x y)(1 - \frac{x}{y})(x - s)(x - \frac{1}{s})} \,,
\label{eq:opp}
\ee
where $s$ is the gauge fugacity $e^{i\alpha}$, and the $O(1)_-$ part contributes
\be
Z^{-}_{1} = \frac{x^2}{(1 - x y)(1 - \frac{x}{y})(x + s)(x + \frac{1}{s})} \,.
 \label{eq:omp}
\ee
The latter is equivalent to inserting a parity twist $(-1)^P$ in the trace defining the index,
where $P$ is the element $-1 \in O(1)$.
The only parity odd modes come from the gauge moduli, hence the sign flip in front of $s$.
The full instanton index is given by combining the two parts, or equivalently by
parity-projecting the spectrum. But here one has a choice of projecting by either
$\frac{1}{2}(1+(-1)^P)$ or $\frac{1}{2}(1-(-1)^P)$.
This is precisely the choice of the $\theta$ parameter, which gives a phase $e^{i\theta}$ 
in the contribution with the Wilson line.
The even-parity projection gives
\be
Z_1^{\theta = 0} = \frac{1}{2}\left(Z_1^+ + Z_1^-\right) = \frac{x^2 (1+x^2)}{(1 - x y)(1 - \frac{x}{y})(1 - (x s)^2)(1 - (\frac{x}{s})^2)} \,,
\label{eq:ipf_0}
\ee  
which is the result of \cite{KKL} for pure $SU(2)$.
For $\theta = \pi$ we take the difference of (\ref{eq:opp}) and (\ref{eq:omp}), which gives
\be
Z^{\theta = \pi}_{1} =  \pm \frac{1}{2}\left(Z_1^+ - Z_1^-\right) = \pm \frac{x^3 (s+\frac{1}{s})}{(1 - x y)
(1 - \frac{x}{y})(1 - (x s)^2)(1 - (\frac{x}{s})^2)} \,.
\label{eq:ipf_pi}
\ee  
We have included a possible sign ambiguity, due to a potential shift in the fermion number
relative to the $\theta = 0$ case.
 We are not able to fix the sign from first principles, but we will give an indirect argument that
the minus sign is the correct one.

\begin{figure}[htbp]
\begin{center}
\includegraphics[scale=0.50]{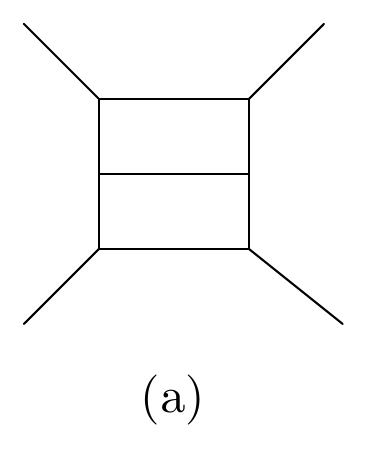}
\hskip 1cm
\includegraphics[scale=0.50]{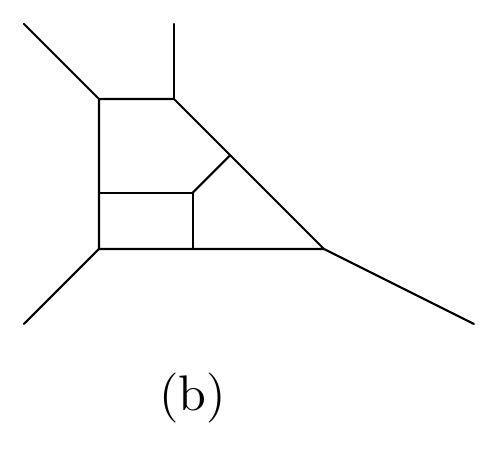}
\caption{The instanton particle: (a) in the $E_1$ theory, (b) in the $\tilde{E}_1$ theory.}
\label{Instanton_particle}
\end{center}
\end{figure}

It is illuminating to expand the result in powers of $x$, since this orders the contributions
roughly according to their scaling dimension.
The leading order contribution corresponds to the ground state of the instanton particle.
We see that to leading order, $Z_1^{\theta = 0} \sim x^2$, whereas
$Z_1^{\theta = \pi} \sim x^3(s + \frac{1}{s})$.
This shows that the ground state of the instanton particle is charged under the gauge symmetry
in the $\tilde{E}_1$ theory, and is gauge-neutral in the $E_1$ theory.
This is something we can also see from the description of BPS states as string-webs inside the 5-brane-web \cite{Aharony:1997bh}, 
Fig.~\ref{Instanton_particle}.
On the Coulomb branch of the $E_1$ theory, the instanton particle is described by a D-string 
between the two NS5-branes.
It is therefore uncharged with respect to the $SU(2)$ gauge symmetry at the origin.
(Although it becomes charged under the unbroken $U(1)$ on the Coulomb branch
due to the one-loop CS term).
In the $\tilde{E}_1$ theory on the other hand, the instanton particle is described a 3-string web
with a fundamental string prong ending on one of the D5-branes.
This corresponds to an $SU(2)$ charge in the fundamental representation,
which is precisely what we see in the leading term in the index.

Let us now return to the question of the overall sign in (\ref{eq:ipf_pi}).
This is related to the question of whether the ground state is bosonic or fermionic,
and as we will now argue, the ground state of the instanton particle in the $\tilde{E}_1$
theory is most likely fermionic.
The basic idea is to examine the symmetry properties of the 2-instanton operator,
and to compare it with the symmetry properties of the product of two 1-instanton operators.
For free particles, the two particle state should be a symmetrized product if they are bosons
and an antisymmetrized product if they are fermions.
This should be seen by comparing the 2-instanton partition function with the symmetrized 
or antisymmetrized product of the 1-instanton partition function.
The instantons are not free particles, so we will not get an equality. 
Nevertheless, the leading terms in $x$ are dominated by the symmetrized (or antisymmetrized) product of the 1-instanton 
moduli space, and so we should be able to determine which possibility of the two is more likely.
The symmetry properties of the products are encoded in the Plethystic exponential.
Extracting the 2-instanton term, in the bosonic case we would get
\be
PE \left.\left[\frac{q x^3 (s+\frac{1}{s})}{(1 - x y)(1 - \frac{x}{y})(1 - (x s)^2)(1 - (\frac{x}{s})^2)}\right] \right|_{q^2} \approx (1+s^2+\frac{1}{s^2})x^6 + O(x^7) \,,
\ee
whereas in the fermionic case we would get
\be
PE \left.\left[-\frac{q x^3 (s+\frac{1}{s})}{(1 - x y)(1 - \frac{x}{y})(1 - (x s)^2)(1 - (\frac{x}{s})^2)}\right]\right|_{q^2} \approx x^6 + O(x^7) \,.
\ee
Comparing with the expansion of the 2-instanton index,
\be
Z_{2}^{\theta = \pi} \approx x^6 + O(x^7) \,,
\ee
we conclude that the ground state of the instanton in the $\tilde{E}_1$ theory is actually fermionic,
and therefore that the correct sign in (\ref{eq:ipf_pi}) is the minus sign.

We can now compute the superconformal index using (\ref{eq:index}).
One can deduce from \cite{KKL} that for small $x$, $Z_k^\pm \sim {\cal O}(x^{2k})$.
So to compute the index to  ${\cal O}(x^{2k})$, we generically need to include instanton contributions 
up instanton number $k$. 
For the $\theta = 0$ theory the authors of \cite{KKL} found\footnote{The result is expressed in terms of $SU(2)$
characters. For example $\chi_{\bf 2}(x) = \sqrt{x} + \frac{1}{\sqrt{x}}$, $\chi_{\bf 3}(x) = x + 1 + \frac{1}{x}$.}
\bea
\label{index_0}
I^{\theta = 0} & = & 1 + \chi_{\bf 3}(q) x^2 + \chi_{\bf 2}(y^2)[1+\chi_{\bf 3}(q)] x^3 
+ \Big(\chi_{\bf 3}(y^2)\big[1+\chi_{\bf 3}(q)\big] + 1 + \chi_{\bf 5}(q)\Big) x^4 \nonumber \\
&+& \Big(\chi_{\bf 4}(y^2)\big[1+\chi_{\bf 3}(q)\big] + \chi_{\bf 2}(y^2)
\big[1+\chi_{\bf 3}(q)+\chi_{\bf 5}(q)\big]\Big) x^5 \nonumber \\
& + & \Big(\chi_{\bf 5}(y^2)\big[1+\chi_{\bf 3}(q)\big] + \chi_{\bf 3}(y^2)
\big[1+\chi_{\bf 3}(q) + \chi_{\bf 5}(q) + \chi_{\bf 3}^2(q)\big]
+ \chi_{\bf 3}(q) + \chi_{\bf 7}(q) - 1\Big)x^6 \nonumber \\
&+& \Big(\chi_{\bf 6}(y^2)\big[1 + \chi_{\bf 3}(q)\big]
+ \chi_{\bf 4}(y^2)\big[2 + 4\chi_{\bf 3}(q) + 2\chi_{\bf 5}(q)\big] \nonumber \\
&& \qquad\qquad \mbox{} + \chi_{\bf 2}(y^2)\big[1 + 3\chi_{\bf 3}(q) + 2\chi_{\bf 5}(q) + \chi_{\bf 7}(q)\big]\Big) x^7 \nonumber \\ 
 & + & \Big(\chi_{\bf 7}(y^2)\big[1 + \chi_{\bf 3}(q)\big] 
 + \chi_{\bf 5}(y^2)\big[3\chi_{\bf 5}(q) + 5\chi_{\bf 3}(q) + 4\big]\nonumber \\
 && \mbox{} + \chi_{\bf 3}(y^2)\big[2\chi_{\bf 7}(q) + 3\chi_{\bf 5}(q) + 7\chi_{\bf 3}(q) + 2\big]
 + \chi_{\bf 9}(q) + 2\chi_{\bf 5}(q) + 2\chi_{\bf 3}(q) + 3\Big) x^8 \nonumber \\
 &+& {\cal O}(x^9) \,.
\eea
This exhibits manifestly the enhancement of the global $U(1)_T$ symmetry to $E_1 = SU(2)$.
In particular the conserved current multiplets contribute to the coefficient of $x^2$,
which is $\chi_{\bf 3}(q) = 1 + q + 1/q$. The three contributions come respectively 
from the perturbative $U(1)_T$ current and two charged currents corresponding to
an instanton particle (D0-brane) and an anti-instanton particle (anti-D0-brane).
For the $\theta = \pi$ theory we find 
\bea
\label{index_pi}
I^{\theta = \pi} & = & 
1 + x^2 + 2 \chi_{\bf 2}(y^2) x^3 +  \Big(1+2\chi_{\bf 3}(y^2)\Big) x^4 +  \Big(2\chi_{\bf 4}(y^2)+\chi_{\bf 2}(y^2)\Big) x^5 \nonumber \\
 & + &  \Big(2\chi_{\bf 5}(y^2) + 3\chi_{\bf 3}(y^2) + q^2 + \frac{1}{q^2}\Big) x^6 \nonumber \\
 &+& \left(2\chi_{\bf 6}(y^2) + 4\chi_{\bf 4}(y^2) + 4\chi_{\bf 2}(y^2) + \chi_{\bf 2}(y^2)(q^2 + \frac{1}{q^2})\right) x^7 \nonumber \\
&+&  \left( 2\chi_{\bf 7}(y^2) + 7\chi_{\bf 5}(y^2) + 7\chi_{\bf 3}(y^2) + 4 +   \chi_{\bf 3}(y^2)(q^2 
+ \frac{1}{q^2})\right) 
x^8 \nonumber \\ 
 &+&  O(x^9) \,.
\eea
As anticipated, there is no symmetry enhancement in this theory.
In fact, it is clear from (\ref{eq:ipf_pi}) that the instanton will only start to contribute at ${\cal O}(x^6)$
in this case.
Our result also     matches perfectly the computation of the index of the $\tilde{E}_1$
theory done in \cite{Iqbal:2012xm}, to the order it was done there.

\subsection{Adding flavor}

Flavor hypermultiplets contribute fermionic zero modes that give factors of the form $(e^{im_i/2} \mp e^{-im_i/2})$ in the numerator of the instanton
index \cite{KKL}, where $m_i$ are the chemical potentials associated to the flavor symmetries (these can be thought of
as the masses of the flavor hypermultiplets).
The sign is correlated with the sign of $O(k)_\pm$.
For $O(k)_+$ the sign is negative due to the $(-1)^F$ operation,
and for $O(k)_-$, corresponding to the insertion of $(-1)^P$, the sign is positive since 
the flavors are parity-odd.
For one flavor, the 1-instanton functions are given by
\be
Z^{\pm}_{1} = \frac{x^2
(e^{im/2} \mp e^{-im/2})}{(1 - x y)(1 - \frac{x}{y})(x \mp s)(x \mp \frac{1}{s})} \,.
\ee
For the two choices of $\theta$ we then get:
\bea
Z^{\theta=0}_{1} &=& \frac{x^2 \left(e^{im/2}(1+x^2)
- x e^{-im/2}(s + \frac{1}{s})\right)}{(1 - x y)(1 - \frac{x}{y})(1 - (x s)^2)(1 - (\frac{x}{s})^2)} \label{eq:ipfwf}\,, \\[5pt]
Z^{\theta=\pi}_1 &=& \frac{x^2 \left(e^{-im/2}(1+x^2)
- x e^{im/2}(s + \frac{1}{s})\right)}{(1 - x y)(1 - \frac{x}{y})(1 - (x s)^2)(1 - (\frac{x}{s})^2)} \,,
\eea
where in the second case we have used the overall minus sign of (\ref{eq:ipf_pi}).
The two indices are related by changing the sign of $m$, and they are equal for $m=0$.
This is as expected, since 
the $\theta$ parameter becomes unphysical in the presence of massless fermions
in the fundamental representation.
This also reinforces our argument for the sign in (\ref{eq:ipf_pi}).


\subsection{An alternative approach}

An alternative way to compute the $SU(2)$ instanton index is to treat the gauge group as $SU(N)$
with $N=2$. Practically speaking, the $SU(N)$ Nekrasov partition function is really for $U(N)$,
but we can freeze-out the overall $U(1)$ by setting its fugacity to 1.
However the distinction is important when we include a CS term.
Although there is no possible CS term for $SU(2)$, one is possible for $U(2)$.
We will see that from the $SU(2)$ point of view this reduces to the discrete $\theta$ parameter.
The Nekrasov 1-instanton partition function for $SU(2)$ with CS level $\kappa$ is given by \cite{KKL}:
\be
Z_1^{SU(2)_\kappa} = \frac{1}{2\pi i} \oint \frac{u^{1+\kappa} (1-x^2) du}
{(1-xy)(1-\frac{x}{y})(u-xs)(u-\frac{x}{s})(u-\frac{s}{x})(u-\frac{1}{xs})} \,,
\ee
where $s$ is the $SU(2)$ gauge fugacity as before. The integral is taken on the unit circle
in the $u$-plane. There are four poles, but if we take $x\ll 1$ only the two at
$u=xs$ and $u=\frac{x}{s}$ contribute, and the result is
\be
Z_1^{SU(2)_\kappa} =  \frac{x^{2+\kappa}\left[(s^{2-\kappa} - s^\kappa)x^2 + s^{2+\kappa} - s^{-\kappa}\right]}
{(1-xy)(1-\frac{x}{y})(1-(xs)^2)(1-(\frac{x}{s})^2)(s^2-1)} \,.
\label{U(2)_instanton_index}
\ee
At CS level $\kappa = 0$ this gives
\be
Z_1^{SU(2)_0} = \frac{x^2 (1+x^2)}{(1 - x y)(1 - \frac{x}{y})(1 - (x s)^2)(1 - (\frac{x}{s})^2)} \,,
\ee
in agreement with the result for the $E_1$ theory, namely the $SU(2)$ theory with $\theta = 0$ (\ref{eq:ipf_0}).
At CS level $\kappa = 1$ we get
\be
Z_1^{SU(2)_1} = \frac{x^3 (s+\frac{1}{s})}{(1 - x y)
(1 - \frac{x}{y})(1 - (x s)^2)(1 - (\frac{x}{s})^2)} \,,
\ee
in agreement, up to a sign, with the $\tilde{E}_1$ theory (\ref{eq:ipf_pi}).

The sign difference above suggests that the procedure of freezing-out the overall 
$U(1)$ by setting its fugacity to 1 leaves a residue of the form $(-1)^\kappa$.
More generally, the result for $Z_k$ with fundamental matter in the $SU(N)$ formalism seems to
have a $U(1)$ residue $(-1)^{k\left(\kappa + N_f/2\right)}$.
We suspect that this is associated with the mixed CS term inherent in the decomposition of $U(2)$
to $U(1)\times SU(2)$,
\be
S_{mixed \, CS} \propto \kappa \int \hat{A} \wedge \mbox{Tr}(F\wedge F) \,.
\ee
The $N_f/2$ contribution reflects the one-loop shift of the CS level due to the flavors.
It would be interesting to understand this better.
 
It is also interesting to 
consider the theory with CS level $\kappa = 2$. The result for the 1-instanton
partition function is given by
\be
Z_1^{SU(2)_2} = \frac{x^4\left[s^2 + 1 + \frac{1}{s^2} - x^2\right]}{(1 - x y)
(1 - \frac{x}{y})(1 - (x s)^2)(1 - (\frac{x}{s})^2)} \,.
\label{SU(2)_2_index}
\ee
However this result is problematic since it is not invariant under $x \rightarrow \frac{1}{x}$,
unlike the results for $\kappa = 0$ and $\kappa = 1$.
This transformation is part of the superconformal group; it's an element of $SU(2)_x\subset SO(5)$.
Therefore it should be respected by the instanton index.
The above result for the index is missing states that are required in order to form 
complete representations of the superconformal group.
We claim that the missing sector can be accounted for by adding a term $\Delta$ to (\ref{SU(2)_2_index}), where
\be
\Delta = \frac{x^2}{(1-xy)(1-\frac{x}{y})} \,.
\ee
The sum is 
\be
Z_1^{SU(2)_2} + \Delta  = \frac{x^2 (1+x^2)}{(1 - x y)(1 - \frac{x}{y})(1 - (x s)^2)(1 - (\frac{x}{s})^2)} \,,
\ee
and is invariant under $x\rightarrow \frac{1}{x}$.
Indeed it is precisely $Z_1^{SU(2)_0}$, namely the 1-instanton index of the $E_1$ theory.
This is what we expect.
The CS level $2$ theory is really the $SU(2)$ theory with $\theta = 2\pi \sim 0$.
It corresponds to the second 5-brane web realization of the $E_1$ theory 
shown in Fig.~\ref{SU(2)_webs}(c). 
More generally, we claim that the full index, including all instanton corrections,
should be multiplied by $PE[q\Delta]$.
A more detailed derivation and interpretation of this result will appear elsewhere \cite{BRZ_1}.
As a teaser, let us however mention that the same term must be added more generally for $SU(N)_N$,
showing that the fixed point theory corresponding to $SU(N)_N$ has an enhanced $SU(2)$ global summetry.

\section{Conclusions}

By properly implementing the effect of the 5d $\theta$ parameter on the instanton particle
in the 5d ${\cal N}=1$ $SU(2)$ theory, we have computed the superconformal index of the $\tilde{E}_1$ theory.
Our result confirms the lack of global symmetry enhancement, and agrees with other approaches.

The generalization to $Sp(N)$ with an antisymmetric hypermultiplet is straightforward, 
and can be obtained in an analogous manner from the results of \cite{KKL}.
The $\theta = \pi$ theory will not have an enhanced global symmetry.
In fact, the instanton contribution starts at a power of $x$ that scales with $N$,
so that at large $N$ the index of the $\tilde{E}_1$ theory becomes purely perturbative 
(for a computation of the large $N$ perturbative index see \cite{Bergman:2013koa}).

It would be interesting to explore further the Type I' string theory description of the discrete $\theta$ parameter,
and in particular its effect on the D0-brane.
Can the difference in the ground states of the instanton particle in the two theories be understood
in terms of some stringy mechanism?

It would also be interesting to understand the supergravity dual of the large $N$ $\tilde{E}_1$ theory,
and related quiver theories, extending the results of \cite{Brandhuber:1999np,Bergman:2012kr}.

\section*{Acknowledgements}

O.B. would like to thank the High Energy Physics group at the University of Oviedo for their hospitality.
O.B. is supported in part by the Israel Science Foundation under grants no. 392/09, and 352/13,
the US-Israel Binational Science Foundation under grants no. 2008-072, and 2012-041,
the German-Israeli Foundation for Scientific Research and Development under grant no. 1156-124.7/2011,
and by the Technion V.P.R Fund.
G.Z. is supported in part by Israel Science Foundation under grant no. 392/09.
D.R-G is supported by the Ram\'on y Cajal fellowship RyC-2011-07593, as well as by the Spansih Ministry of Science and Education grant FPA2012-35043-C02-02.


\begin{thebibliography}{99}


\bibitem{Seiberg:1996bd}
  N.~Seiberg,
  ``Five-dimensional SUSY field theories, nontrivial fixed points and string dynamics,''
  Phys.\ Lett.\ B {\bf 388}, 753 (1996)
  [hep-th/9608111].
  
\bibitem{Morrison:1996xf} 
  D.~R.~Morrison and N.~Seiberg,
  Nucl.\ Phys.\ B {\bf 483}, 229 (1997)
  [hep-th/9609070].
  
\bibitem{Douglas:1996xp} 
  M.~R.~Douglas, S.~H.~Katz and C.~Vafa,
  Nucl.\ Phys.\ B {\bf 497}, 155 (1997)
  [hep-th/9609071].
  
\bibitem{Intriligator:1997pq}
  K.~A.~Intriligator, D.~R.~Morrison and N.~Seiberg,
  ``Five-dimensional supersymmetric gauge theories and degenerations of Calabi-Yau spaces,''
  Nucl.\ Phys.\ B {\bf 497}, 56 (1997)
  [hep-th/9702198].
  

  
\bibitem{Aharony:1997ju} 
  O.~Aharony and A.~Hanany,
  Nucl.\ Phys.\ B {\bf 504}, 239 (1997)
  [hep-th/9704170].
  
 

\bibitem{Polchinski:1995df}
  J.~Polchinski and E.~Witten,
  ``Evidence for heterotic - type I string duality,''
  Nucl.\ Phys.\ B {\bf 460}, 525 (1996)
  [hep-th/9510169].
  
\bibitem{Matalliotakis:1997qe}
  D.~Matalliotakis, H.~-P.~Nilles and S.~Theisen,
  ``Matching the BPS spectra of heterotic Type I and Type I-prime strings,''
  Phys.\ Lett.\ B {\bf 421}, 169 (1998)
  [hep-th/9710247].

\bibitem{Bergman:1997py}
  O.~Bergman, M.~R.~Gaberdiel and G.~Lifschytz,
  ``String creation and heterotic type I' duality,''
  Nucl.\ Phys.\ B {\bf 524}, 524 (1998)
  [hep-th/9711098].




\bibitem{KKL}
  H.~-C.~Kim, S.~-S.~Kim and K.~Lee,
  ``5-dim Superconformal Index with Enhanced En Global Symmetry,''
  JHEP {\bf 1210}, 142 (2012)
  [arXiv:1206.6781 [hep-th]].

\bibitem{Witten:1982fp} 
  E.~Witten,
  Phys.\ Lett.\ B {\bf 117}, 324 (1982).

  
   
  
  
   
  
\bibitem{Sethi:2013hra} 
  S.~Sethi,
  arXiv:1304.1551 [hep-th].

\bibitem{Bergman:1999ta} 
  O.~Bergman, E.~G.~Gimon and P.~Horava,
  JHEP {\bf 9904}, 010 (1999)
  [hep-th/9902160].
  
   
\bibitem{Sen:1998ki} 
  A.~Sen,
  JHEP {\bf 9810}, 021 (1998)
  [hep-th/9809111].

  
\bibitem{Witten:1998cd} 
  E.~Witten,
  JHEP {\bf 9812}, 019 (1998)
  [hep-th/9810188].
  
\bibitem{Sen:1998tt} 
  A.~Sen,
  JHEP {\bf 9809}, 023 (1998)
  [hep-th/9808141].

  
\bibitem{Kinney:2005ej}
  J.~Kinney, J.~M.~Maldacena, S.~Minwalla and S.~Raju,
  Commun.\ Math.\ Phys.\  {\bf 275}, 209 (2007)
  [hep-th/0510251].
  
\bibitem{Nekrasov:2002qd} 
  N.~A.~Nekrasov,
  Adv.\ Theor.\ Math.\ Phys.\  {\bf 7}, 831 (2004)
  [hep-th/0206161].
  
\bibitem{Nekrasov:2004vw} 
  N.~Nekrasov and S.~Shadchin,
  Commun.\ Math.\ Phys.\  {\bf 252}, 359 (2004)
  [hep-th/0404225].

\bibitem{Aharony:1997bh} 
  O.~Aharony, A.~Hanany and B.~Kol,
  JHEP {\bf 9801}, 002 (1998)
  [hep-th/9710116].
  
\bibitem{Iqbal:2012xm} 
  A.~Iqbal and C.~Vafa,
  arXiv:1210.3605 [hep-th].
  


  
  \bibitem{BRZ_1}
  O.~Bergman, D.~Rodriguez-Gomez, G.~Zafrir, to appear.
  

\bibitem{Bergman:2013koa} 
  O.~Bergman, D.~Rodriguez-Gomez and G.~Zafrir,
  JHEP {\bf 1308}, 081 (2013)
  [arXiv:1305.6870 [hep-th]].


\bibitem{Brandhuber:1999np}
  A.~Brandhuber and Y.~Oz,
  ``The D-4 - D-8 brane system and five-dimensional fixed points,''
  Phys.\ Lett.\ B {\bf 460}, 307 (1999)
  [hep-th/9905148].



\bibitem{Bergman:2012kr}
  O.~Bergman and D.~Rodriguez-Gomez,
  ``5d quivers and their AdS(6) duals,''
  JHEP {\bf 1207}, 171 (2012)
  [arXiv:1206.3503 [hep-th]].












\end{thebibliography}
\end{document}